# Mining Crash Fix Patterns

Jaechang Nam and Ning Chen
Department of Computer Science and Engineering
The Hong Kong University of Science and Technology
China
{jcnam,ning}@cse.ust.hk

## ABSTRACT

During the life cycle of software development, developers have to fix different kinds of bugs reported by testers or end users. The efficiency and effectiveness of fixing bugs have a huge impact on the reliability of the software as well as the productivity of the development team. Software companies usually spend a large amount of money and human resources on the testing and bug fixing departments. As a result, a better and more reliable way to fix bugs is highly desired by them. In order to achieve such goal, in depth studies on the characteristics of bug fixes from well maintained, highly popular software projects are necessary.

In this paper, we study the bug fixing histories extracted from the Eclipse project, a well maintained, highly popular open source project. After analyzing more than 36,000 bugs that belongs to three major kinds of exception types, we are able to reveal some common fix types that are frequently used to fix certain kinds of program exceptions. Our analysis shows that almost all of the exceptions that belong to a certain exception can be fixed by less than ten fix types. Our result implies that most of the bugs in software projects can be and should be fixed by only a few common fix patterns.

## 1. INTRODUCTION

As software programs evolve over time, more and more bugs are introduced and later fixed. Software developers usually have to spend a large portion of their time trying to fix various bugs reported by the users. Therefore, the increase of efficiency on fixing bugs can highly promote software productivity. To increase the efficiency on fixing bugs, we need to have in depth knowledge about how bugs were fixed in the past.

In this paper, we focus our study on how developers fix various kinds of program bugs. Even though different kinds of bugs have different characteristics and may require totally different ways to fix, there can be some hidden common patterns that developers use to fix the same category of bugs. In our study, we choose the Eclipse project [1] as the subject of study. We retrieve its version control information from the Kenyon database, and extract all of the commits that are related to fixing some bugs [2]. To further improve our data, we also use the Bird data which links code commits with bug reports manually [3]. We carefully study that information and try to find out if there exists any hidden common fix patterns for a particular type of crashes. We also apply in depth analysis on the fix patterns and try to explain why they are commonly used to fix the particular type of crashes.

This paper makes the following major contributions:

- Understand common crash fix patterns for major software crashes.

- Provide preliminary results to identify crash fix patterns.

- Suggest possible applications based on crash fix pattern mining results.

The rest of the paper is structured as follows: Section 2 presents motivation and related work of this study. Section 3 describes the detailed methodology to mine fix patterns, and Section 4 and 5 present the result of fix pattern mining and its analysis respectively. Finally, section 6 concludes our study on fix pattern mining.

## 2. RELATED WORK

Previous studies such as GrouMiner [7] have focused a lot on the common patterns of the program codes, i.e. API usage patterns. They are mainly interested in mining the common patterns of individual versions of code. Different from the previous works, we focus our study on the delta between buggy code and fixed code, and try to find out the common patterns of the bug fixes for different kinds of exceptions. Our hypothesis is that, developers tend to fix the same kind of exceptions, such as the null pointer exception or the index out of bounds exception in only a few common patterns. By studying the bug fixing commits from the Eclipse project, we try to find out the common approaches by the developers to resolve different kinds of exceptions.

Sudhakrishnan et al. analyzed bug fix history of four hardware projects written in Verilog and revealed 25 bug fix patterns [10]. They also found out that most bug fixes fall into a few fix patterns.

Livshits and Zimmermann proposed DynaMine, a tool that combines revision history mining and dynamic analysis techniques to discover new application-specific patterns [6]. They have discovered 56 previously unknown, highly application specific-patterns with their tool.

Weißgerber et al. study the common characteristics of the accepted patches [11]. They found that more than half of the submitted patches from their subject projects change only one or two lines. Moreover, they discover that small patches have a much higher acceptance rate than those that change a lot of lines.

Fluri et al. proposed an approach to discover patterns of change types [4]. They observed that change type patterns contain development activity information and can affect the control flow.

BugMem, a bug finding tool, implemented by Kim et al. [5] mined software change histories from software repositories. Based on this tool, project-specific bugs are detected and also corresponding fixes are suggested by finding bug and fix pairs. As a result, they insisted the possibility of providing a strong suggestion for the fix. However, the tool has also a high false positive rate about the changes of non-bugs. To realize the strong suggestion for the fix, we believe that analyzing bug fix patterns are necessary.

Pan et al. [8] identified 27 bug fix patterns from software project change histories of seven Java open source projects, Eclipse,

Columba, JEdit, Scarab, ArgoUML, Lucene, and MegaMek. As the 27 bug fix patterns provide a promising guide to classify general bug fix patterns, the results are useful to understand the bug fix patterns of software. However, they just focused on syntactical structures of fix patterns and did not consider the implication of bug fix patterns. In other words, the issue regarding which fix patterns are related which bugs are not investigated much.

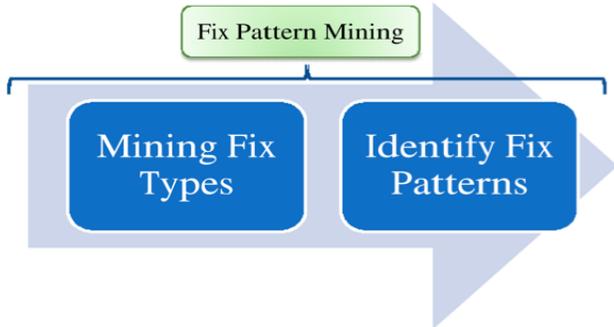

**Figure 1. An approach of fix pattern mining.**

If we can identify the common fix patterns relating to specific crashes, it can be useful in various ways. First of all, it provides developers with concrete information on how previous similar bugs were fixed and guide developers to fix the current bugs effectively. Although, this kind of information seems trivial in some situations for experienced developers, it is especially useful for developers who are new to the project and have no knowledge on how bugs were fixed within the project in the past by other experienced developers. Secondly, by leveraging the fix pattern information, it is possible to detect anomaly fixes. Although developers who fix the bugs already have knowledge on why and how the program code fails, it is still possible that the fixes they committed introduce new bugs or do not fix the bugs completely. This is especially common for inexperienced developers who lack knowledge on how similar bugs were fixed in the past. By comparing the committed fixes with the fix patterns, we can reveal potential failure prone fixes and give warnings to the developers before they actually commit their codes to the code repository. Thirdly, fix patterns information might be able to help software to automatically fix themselves after a crash. As shown by our study, the major kinds of exceptions can be fixed with a few common patterns. As a result, it is possible to develop heuristic algorithms that can fix a crash program automatically according to the type of crash and its context when crashed.

**Table 1. Statistical numbers of Kenyon and the Bird data**

| Total Bugs | 36,626 |
|---|---|
| Total File Revisions | 64,137 |
| NullPointerException (NPE) | 1,575 |
| OutOfBoundsException (OOBE) | 243 |
| ClassCastException (CCE) | 181 |

## 3. METHODOLOGY

In this section, we present how to mine software repository to get crash fix patterns. Figure 1 shows the approach of fix pattern mining. Simply, we can divide it into two steps. The first step is to get fix types that are recurring fixes to solve software crashes. The second step is to identify fix patterns based on the fix types of the first step. In this report, we focused on the first step that contains two sub steps: identify major crashes, and collect fix types. All these steps are applied to the Eclipse software repository that includes Software Configuration Management (SCM) system, concurrent versions system (CVS), and bug tracking system, BugZilla.

### 3.1 Identify Major Crashes

To get the major crashes of the Eclipse, we used data extracted by using Kenyon framework, which is a feature extractor for software repositories [2]. One issue to get the major crashes is that CVS and BugZilla were separated and it is not easy to get information about which bugs are fixed in which revisions. The Bird data is a possible solution to overcome this issue. However, even the Bird data still has bias issues [3]. Basically, the Bird data contains linkage information between CVS and BugZilla. The linkage information is generated by both automatic and manual ways. Thus, this represents the corresponding fix of each bug. Some statistic results from Kenyon and the Bird data are shown in Table 1. To find crashes, we focus on Java exceptions, which make programs abnormally stop.

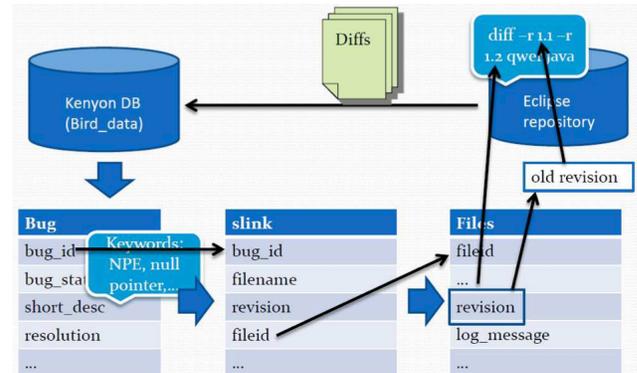

**Figure 2. Collect Fix Types**

### 3.2 Collect fix types

Figure 2 shows how to collect fix types from the Bird data. The required tables to get fix types are 'bug', 'slink', and 'files' from Kenyon database. The table 'bug' contains the records of bug reports and we select a set of 'bug id' corresponding to each Java exception by using keywords mapped in the field 'short desc'. Based on the set of 'bug id', we can get 'fileid' from the table 'slink' and the 'fileid' is used to get the revision numbers of the fix of the bug reports. From these revision numbers, we can compute the old revision numbers, which are used to get the differences (diffs) between bug and fix files. The diff results are stored into the database again and we analyze them to get fix types because the diff results are the actual fixes to solve crashes.

## 4. RESULT

This section summarizes results coming from software repository by using the methodologies presented in the prevision section.

### 4.1 Major crashes

Figure 3 shows the major crashes of the Eclipse project. As we discussed in the previous section, these crashes were extracted from bug reports with keywords such as 'NullPointer', 'NPE', 'CCE', 'OOBE', and other related keywords. To remove the wrong records, we manually checked the selected records. The top crash is the NullPointerException (79%) followed by IndexOutOfBoundException (12%). The third crash is

ClassCastException (9%). Based on this result, we investigate these three crashes in detail.

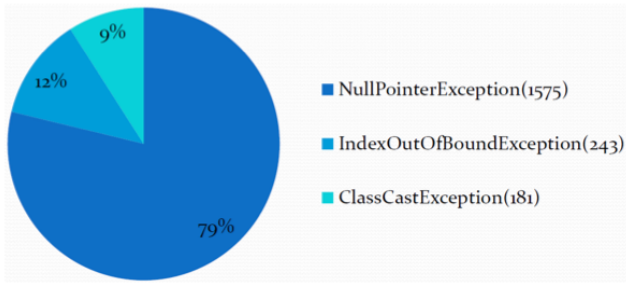

Figure 3. Major crashes of the Eclipse project.

## 4.2 Relationship between the number of file changes and fixes

The number of file changes of a fix implies important information. That is how simple finding fix types is. In other words, if programmers just need to change one file to fix a crash, it means identifying the fix type of a crash is simpler than the case which has to change more files to fix the crash. This fact also means the complex fix make the fix as a generalized fix type. In this sense, Figure 4, 5, and 6 show promising results. To fix a crash, it is required to fix one or two (more than 90%) files as shown in Figures.

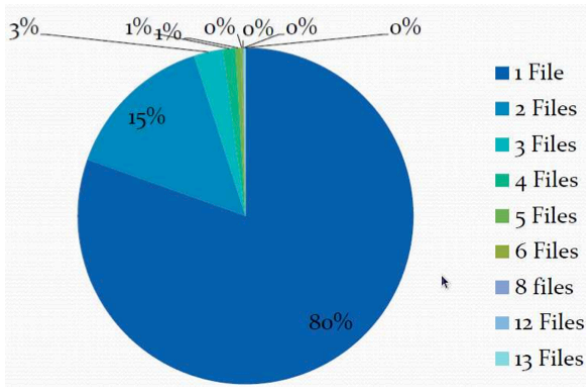

Figure 4. The file changes of a NullPointerException fix.

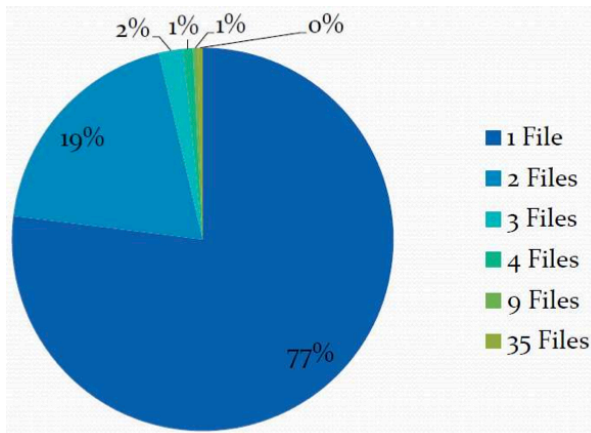

Figure 5. The file changes of IndexOutOfBoundException fix.

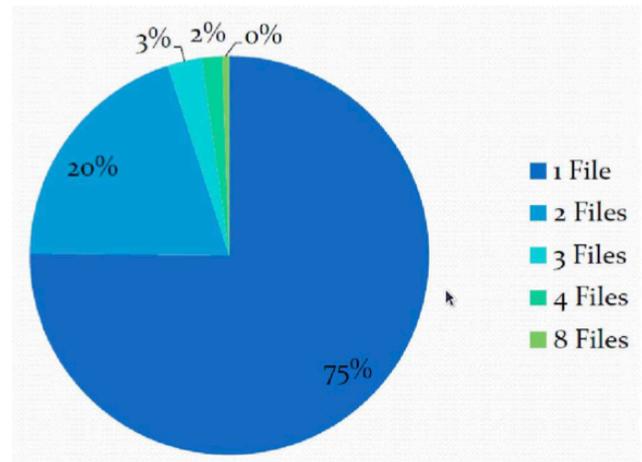

Figure 6. The file changes of a ClassCastException fix.

## 4.3 Fix types

For the NullPointerException, we identified five fix types. NullPointerException occurs when a null object is accessed. Thus, this exception can be most likely prevented by checking whether an object which should be accessed is null or not (null checking). In fact, a major portion of fix types of NullPointerException is 'Null Checker' shown in Figure 7. Other fix types are the indirect ways to avoid null objects as well.

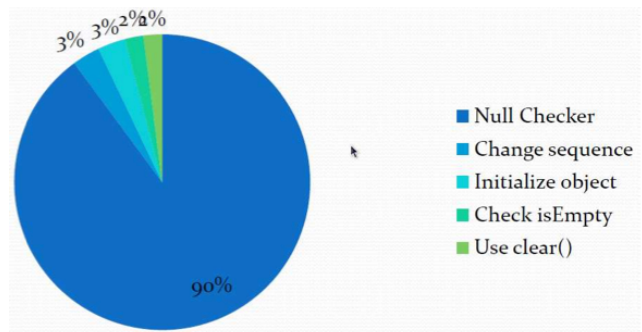

Figure 7. The fix types of NullPointerException.

Figure 8 shows the fix types of IndexOutOfBoundException. This crash is caused when the non-existing array index is used. Namely, most of fixes are related with the length or range of array. The three major fix types are Check Array Length, Check Range, and Fix Off-by-one Error respectively. These three fix types form 72% of all fix types.

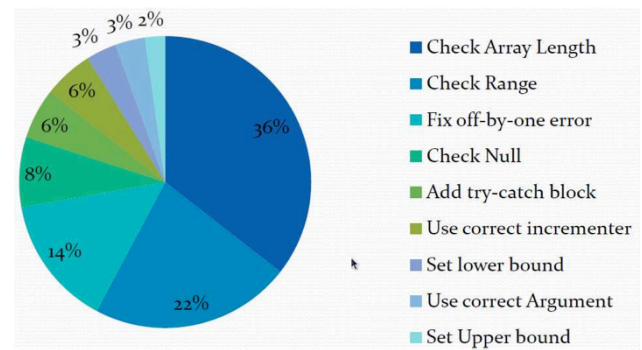

Figure 8. Fix types of IndexOutOfBoundException.

The six fix types identified for ClassCastException are Use instance checker, Use Super Type, Lazy Cast, Use Correct Castee, Type Checking, and Use Correct Caster as shown in Figure 9. The ClassCastException is caused by the following reasons: use of wrong castee or caster. The first five fix types solve the wrong use of castee. The first three fix types cover about 80% of all fix types. Most of ClassCastException were fixed by using instance checker (59%).

All examples of fix types are explained in the appendix.

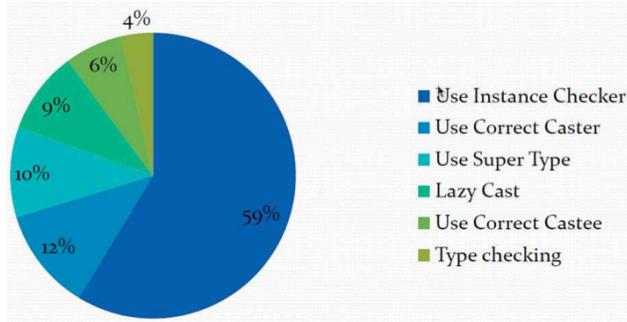

**Figure 9. Fix types of ClassCastException.**

## 5. ANALYSIS

In our study, a total number of 1,999 fixes from both the Kenyon database and the Bird dataset are examined. About 80% of the fixes require only one file modification. About 16% and 3% of the fixes require two and three file modifications. Generally speaking, developers only need to touch a very small subset of files in order to fix a particular crash. This makes identifying fix patterns simpler and facilitate automatic collecting of crash fix types from other subject projects. For every major kind of crashes, there are only a few types of patterns that are common in almost all of the fixes. For instance, in order to fix a NullPointerException, most of the fixes can be categorized as simply adding a null checker, changing statement sequences, and initializing an object. 96% of all the fixes to NullPointerException belong to one of the three patterns. For IndexOutOfBoundException, top three fixes are checking array length / range and fixing off-by-one error. These three fix types can cover 72% of all fixes to this exception. And in order to fix a ClassCastException, most of the fixes are involved with adding "instanceof" statement, change the caster, or use a super type. 81% of all the fixes to a ClassCastException belong to one of the three fix types.

### 5.1 Possible applications

Although the current results of collected fix types are preliminary, they contain meaningful implications as we discussed. As a result, we can think of several possible applications based on crash fix types. If we can identify concrete fix patterns with their corresponding buggy contexts, it would be possible to recommend a proper fix for a certain crash. In other words, when programmers who just involve into the software projects face software crashes and have to fix them, recommending possible fixes to them will be helpful. As research about self-defending software is growing [9], fix patterns can be applied to generate patches automatically. Of course, to realize this idea, we have to generate a precise patch from the fix pattern of a specific defect of software. However, fix patterns will be useful data for empirical study to develop self-defending software. Another possible application is bug identification. Research on bug repository has taken benefits from software repository mining. However, it is difficult to get all information by mining software repositories. Namely, incomplete information can make interpretation bias [3]. In this context, fix patterns can be used to decide whether a change of source code is a fix of a certain defect or not.

## 6. CONCLUSION

We presented how to mine crash fix types. Actually, the current result is too preliminary to generalize each fix pattern. In addition to this, all fix types are collected manually. However, during the process, we have noticed important implications of this study. First, all fix types of NullPointerException, IndexOutOfBoundException, and ClassCastException extracted from Eclipse project could be collected from a few files of revisions containing crash fixes. This fact implies that the collected fix types clearly represents the file changes that are deeply related with the fixes with less noise. So, we believe that it facilitates identifying fix patterns. Second, top three fix types cover about 81–96% of fixes in all exceptions we analyzed. This shows identified fix patterns can be available in various applications with high reliability. On the other hand, the small number of fix types can reduce flexibility to a variety of bugs or crashes. Because all fix types were collected only from the Eclipse project.

Through this study, we identified three major types of exceptions from the Eclipse project. We collect five fix types of NullPointerException and 'Null checker' is the top fix type covering 90% of all fix types. For IndexOutOfBoundException, nine fix types are collected and the three fix types, 'Check array length', 'Check range', and 'Fix off-by-one error', cover 72% of all. In the case of ClassCastException, six fix types are collected and the fix type, 'Use Instance Checker', covers about 60% out of all fixes, followed by 'Use Correct Caster' (12%). According to this result, we see the fact that there are fix patters for major crashes, this fact will drive the further work to identify fix patterns. For the next step, we will investigate the relationship between the fix types and their context in the source code to identify crash fix patterns.

# APPENDIX
## A. FIX TYPE EXAMPLE
### A.1 Fix types of NullPointerExeption
Most of fix types is to prevent null status of objects directly or indirectly.

#### A.1.1 Null Checker
Null checker is a common fix for NullPointerExeption. Programmers just need to add a condition to check whether an object is null or not. After checking null, a fix has different subsequent executions such as return null, continue in a loop, and etc.

- Buggy code
```
public ITextHover getCurrentTextHover(){
   return fTextHoverManager.getCurrentTextHover();
}
```
- Fixed code
```
public ITextHover getCurrentTextHover() {
   if (fTextHoverManager== null)
      return null;
   return fTextHoverManager.getCurrentTextHover();
}
```

#### A.1.2 Change Sequence
To fix NullPointerExeption, the order of statements is changed.
- Buggy code
```
protected void handleNextSelectedNode(ASTNodenode){
   checkParent(node);
   super.handleNextSelectedNode(node);
}
```
- Fixed code
```
protected void handleNextSelectedNode(ASTNodenode){
   super.handleNextSelectedNode(node);
   checkParent(node);
}
```

#### A.1.3 Initialize Object
- Buggy code
```
if (buildTime== null) {
   fProjectBuildTimes.put(project, new ProjectBuildTime());
}
buildTime.setCurrentBuildDate(currentDate);
```
- Fixed code
```
if (buildTime== null) {
   buildTime= new ProjectBuildTime();
   fProjectBuildTimes.put(project, buildTime);
}
buildTime.setCurrentBuildDate(currentDate);
```

#### A.1.4 Use isEmpty()
- Buggy code
```
if (obj instanceof IAdaptable) {
   IAdaptable element = (IAdaptable)obj;
   if (verifyElement(element)==false) return false;
}
```
- Fixed code
```
if (classes.isEmpty()) return true;
if (obj instanceof IAdaptable) {
   IAdaptable element = (IAdaptable)obj;
   If (verifyElement(element)==false) return false;
}
```

#### A.1.5 Initialize object
- Buggy code
```
if (buildTime== null) {
   fProjectBuildTimes.put(project,  new ProjectBuildTime());
}
buildTime.setCurrentBuildDate(currentDate);
```
- Fixed code
```
if (buildTime== null) {
   buildTime= new ProjectBuildTime();
   fProjectBuildTimes.put(project, buildTime);
}
buildTime.setCurrentBuildDate(currentDate);
```

### A.1.6 Use clear()
To reuse data structures such as HashMap, they should be initialized by using their own method clear() rather than setting 'null'.

• Buggy code

  stringToFont= null;

  listeners = null;

• Fixed code

  stringToFont.clear();

  listeners.clear();

## A.2 Fix types of IndexOutOfBoundException
IndexOutOfBoundException is caused by the wrong access of array index. Thus, to fix this kind of defect, controlling and managing the length of range of array is important.

### A.2.1 Check array length
• Buggy code

  Object data = selections[selections.length-1].getData();

  IValueval= null;

  if (data instanceof IndexedVariablePartition) {

    // no details for paritions

    return;

  }

• Fixed code

  if (selections.length> 0) {

    Object data =

      selections[selections.length-1].getData();

    IValueval= null;

    if (data instanceof IndexedVariablePartition) {

      // no details for paritions

      return;

    }

### A.2.2 Check range
• Buggy code

  ICompletionProposal current= fFilteredProposals[index];

  item.setText(current.getDisplayString());

  item.setImage(current.getImage());

  item.setData(current);

• Fixed code

  if (0 <= index && index < fFilteredProposals.length){

    ICompletionProposal current= fFilteredProposals[index];

    item.setText(current.getDisplayString());

    item.setImage(current.getImage());

    item.setData(current);

  }

### A.2.3 Fix off-by-one error
Off-by-one error is a common mistake by programmers confusing the start index of array or the issue whether including the equal sign '=' in the condition with '<' or '>'.

• Buggy code

  error = error.getChildren()[1];

• Fixed code

  error = error.getChildren()[0];

### A.2.4 Add try-catch block
• Buggy code

  IJobChangeListener listener =

        (IJobChangeListener) global.get(i);

• Fixed code

  IJobChangeListener listener = null;

  try {

    listener =

      (IJobChangeListener) global.get(i);

  } catch (ArrayIndexOutOfBoundsExceptione) { }

### A.2.5 Use Correct Increment
• Buggy code

  for (inti=0; i< contentTypes.length; i++) {

    array =registry.getEditorsForContentType(contentTypes[i]);

    for (intj = 0; j < array.length; j++) {

      IEditorDescriptor editor = array[i];

• Fixed code

  for (inti= 0; i< contentTypes.length; i++) {

    array = registry.getEditorsForContentType(contentTypes[i]);

    for (intj = 0; j < array.length; j++) {

      IEditorDescriptor editor = array[j];

### A.2.6 Set Lower Bound
• Buggy code

  fTree.clear(fTree.indexOf(item), true);

• Fixed code

  int index = fTree.indexOf(item);

  if (index >= 0)

    fTree.clear(index, true);

### A.2.7 Use Correct Argument
• Buggy code

  if (index < keyStrokesLength) {

    System.arraycopy(keyStrokes, index,

      newKeyStrokes, index + 1, keyStrokesLength);

  }

• Fixed code

  if (index < keyStrokesLength) {

    System.arraycopy(keyStrokes, index,

      newKeyStrokes, index + 1, keyStrokesLength-index);

  }

### A.2.8 Set Upper Bound
• Buggy code

  int index =

    availableWidth/gc.getFontMetrics().getAverageCharWidth();

• Fixed code

  int index = Math.min(availableWidth)/

      gc.getFontMetrics().getAverageCharWidth(),text,length());

## A.3 Fix types of ClassCaseException

ClassCastException easily occurs when an object implemented by multiple interfaces. Thus, checking the object with a specific interface by using 'instanceof' keyword is the direct way to fix this problem. Using correct caster or castee is another direct solution.

*A.3.1 Use instance checker*

• Buggy code

  ICompilationUnit cu= (ICompilationUnit)

      JavaCore.create(resource);

• Fixed code

  IJavaElement element= JavaCore.create(resource);

  if (! (element instanceof ICompilationUnit))

    continue;

  ICompilationUnit cu= (ICompilationUnit)

      JavaCore.create(resource);

*A.3.2 Use instance checker, lazy case*

In the case of ClassCastException, a set of fixes may be used. As shown in the following example of fix code, class casting is conducted after Instance checker.

• Buggy code

  fContainer= (IContainer) root.findMember(fContainerFullPath);

  if (fContainer!= null)

• Fixed code

  IResourcefound= root.findMember(fContainerFullPath);

  if (found instanceof IContainer) {

    fContainer= (IContainer) found;

    if (fContainer!= null)

*A.3.3 Use correct caste*

• Buggy code

  IJavaType type= (IJavaType)pop();

  IJavaObject object= (IJavaObject)pop();

• Fixed code

  IJavaType type= (IJavaType)pop();

  IJavaObject object= (IJavaObject)popValue();

*A.3.4 Use correct caster*

• Buggy code

  WorkspacePluginModelBase workspaceModelBase=

    (WorkspacePluginModelBase)dialog.getFirstResult();

  fPluginIdText.setText(

      workspaceModelBase.getPluginBase().getId());

• Fixed code

  IPluginModelBase workspaceModelBase=

    (IPluginModelBase)dialog.getFirstResult();

  fPluginIdText.setText(

      workspaceModelBase.getPluginBase().getId());